\documentstyle[aps,twocolumn,eqsecnum,amssymb]{revtex}
\newcommand{\beq}{\begin{equation}}
\newcommand{\eeq}{\end{equation}}
\newcommand{\md}{{\rm d}}
\newcommand{\e}{{\rm e}}
\newcommand{\imu}{{\rm i}}
\newcommand{\eq}[1]{Eq.~(\ref{#1})}
\newcommand{\mrm}[1]{\mathrm{#1}}
\newcommand{\mcal}[1]{{\mathcal #1}}
\newcommand{\re}{\mathop{\rm Re}}
\newcommand{\im}{\mathop{\rm Im}}
\newcommand{\vspan}{\mathop{\rm span}}
\begin{document}
\draft
\title{The Interpretation of Quantum-Mechanical Models with 
Non-Hermitian Hamiltonians and Real Spectra}
\author{R. Kretschmer\thanks{E-mail: 
kretschm@hepth2.physik.uni-siegen.de}}
\address{Fachbereich Physik, Universit\"at Siegen, Germany}
\author{L. Szymanowski\thanks{E-mail:
lech.szymanowski@fuw.edu.pl}}
\address{Institut f\"ur Theoretische Physik, Universit\"at
Regensburg, Germany,\\
and\\
Soltan Institute for Nuclear Studies, Warsaw, Poland}
\date{11 May 2001}
\maketitle
\begin{abstract}
We study the quantum-mechanical interpretation of models with 
non-Hermitian Hamiltonians and real spectra. We set up a general 
framework for the analysis of such systems in terms of Hermitian 
Hamiltonians defined in the usual Hilbert space 
$L_2(- \infty, \infty)$. Special emphasis is put on the correct 
definition of the algebra of physical observables. Within this scheme 
we consider various examples, including the model recently introduced 
by Cannata et~al.\ and the model of Hatano and Nelson. 
\end{abstract}


\section{Introduction}
\label{s1}

Recently, models with non-Hermitian Hamiltonians (non-Hermiticity 
meant here in the sense of the space $L_2(- \infty, \infty)$) have 
attracted a lot of interest. One of the earliest attempts to use such 
systems is the work of Hatano and Nelson \cite{Hatano...} where it was 
suggested that a delocalization transition in superconductors can be 
described by a non-Hermitian Hamiltonian. Next, Bender and 
collaborators in a series of papers \cite{Bender...} investigated some 
non-Hermitian, $\mcal{PT}$ symmetric Hamiltonians, arguing that 
they have real eigenvalues. This has triggered a lot of activity in 
this field \cite{Znojil1,Bagchi,Andrianov,Fernandez,Znojil2,Roy,%
Bender1,Basu-Malick,Dorey}.

Despite this, some fundamental issues concerning the 
quantum-mechanical interpretation have only started to be addressed
\cite{Bender2,Beckers,Znojil3,Japaridze}. In the models treated by 
Bender et al.\ it is, for example, necessary to extend the definition 
of position-space wave functions to complex values of the coordinate 
\cite{Bender...,Bender3}. This means that the wave functions are not 
elements of the Hilbert space $L_2(- \infty, \infty)$, so that the 
notion of non-Hermiticity in the sense of the space $L_2$ does not 
seem to be useful here. In our opinion, the physical meaning of these 
wave functions deserves additional study.

In this paper we will address these questions. In Section~\ref{s2} we 
start by constructing a Hilbert space $\mcal{H}$ that contains  
superpositions of eigenfunctions with real eigenvalues of a 
Hamiltonian $H$. Here the choice of a scalar product that provides the 
link to a probabilistic interpretation is quite arbitrary. Motivated 
by the approach suggested by Bender and collaborators 
\cite{Bender...}, we consider, e.~g., scalar products that are defined 
along complex paths. 

We remove the arbitrariness for the choice of the scalar product 
by demanding that the Hamiltonian is to be interpreted as the 
generator of the time evolution for a closed system, and therefore has 
to be Hermitian in the space $\mcal{H}$. This still leaves some 
arbitrariness for the structure of $\mcal{H}$, but all allowed 
theories have a consistent quantum-mechanical interpretation. Next we
consider the algebra of physical observables in the theory. We set up
a canonical formulation in which the Hamiltonian is only a function of 
two Hermitian operators $x^{\mrm{c}}$ and $p^{\mrm{c}}$ fulfilling
canonical commutation relations. Such operators may under some 
conditions have an interpretation as a position-space variable and the 
generator of its translations, resp. One can then employ the 
uniqueness theorem by von Neumann \cite{Prugovecki} to map the space 
$\mcal{H}$ to the space $L_2$. In this way, the physical 
interpretation is fixed uniquely.

The whole construction is applied to two exactly solvable examples in 
Section~\ref{s3}. The first example is a simple non-Hermitian model 
based on the well-known, Hermitian one-dimensional Coulomb problem on 
the real half axis. Here we show that the canonical formulation does 
not contain any new information. The second example is based on the 
model introduced by Cannata et al. \cite{Cannata} in which the 
potential is given by $V(x) = \e^{2 \imu x} / 2$. Here we also 
construct a canonical formulation $H = H(x^{\mrm{c}}, p^{\mrm{c}})$ 
with Hermitian $H$, but then we show that this formulation does not 
allow the interpretation of $x^{\mrm{c}}$ and $p^{\mrm{c}}$ as 
position and momentum operators, resp., so that the interpretation of 
the model remains unclear. We end this Section with a discussion of 
the phenomenologically important model of Hatano and Nelson 
\cite{Hatano...} in the spirit of our approach.
 
Our concluding remarks are contained in Section~\ref{s4}. Some 
technical details that are used in Section~\ref{s3} are explained in
Appendix~\ref{sa}. In Appendix~\ref{sb} we briefly comment on the 
connection between real spectra and $\mcal{PT}$ symmetric 
Hamiltonians. 

\section{General framework}
\label{s2}

Suppose we have a non-Hermitian (in the usual sense of the space 
$L_2(- \infty, \infty)$) Hamiltonian $H$ and some eigenfunctions 
$\psi_n(x)$ with real eigenvalues $E_n$,
\beq\label{2i}
H(x, p) \psi_n = E_n \psi_n \quad,\quad E_n \mbox{ real} \quad.
\eeq
Here the operators $x$ and $p$ act in the usual way as multiplication
by $x$ and $- \imu$ times differentiation, resp., on $\psi_n(x)$, and 
the eigenfunctions are calculated as solutions of the corresponding 
differential equation. They are not required to be normalizable with 
respect to the norm $\|\psi\|_{L_2} = \sqrt{(\psi, \psi)_{L_2}}$ of 
the Hilbert space $L_2(- \infty, \infty)$. (We use the notation
$(\psi, \psi')_{L_2} = \int_{- \infty}^\infty \md x \, \psi^*(x)
\psi'(x)$ for the scalar product of $L_2$.)

Our aim is to set up a formulation in which we can interpret the
superpositions of the eigenfunctions $\psi_n$ quantum-mechanically. 
For simplicity, we will do this only for the case of a discrete, 
infinite, non-degenerate spectrum of $H$. 

To start, we define the vector space 
$\mcal{V} = \vspan \{\psi_n, n = 0, \ldots\}$ of the finite 
superpositions of the eigenfunctions. Since eigenfunctions 
corresponding to different eigenvalues are always linearly 
independent, the dimension of $\mcal{V}$ is infinite. Next we define a 
scalar product $(.,.)_\mcal{V}$ on this space. Besides requiring the 
usual properties 
($(\psi, c_1 \varphi_1 + c_2 \varphi_2)_\mcal{V} 
= c_1 (\psi, \varphi_1)_\mcal{V} + c_2 (\psi, \varphi_2)_\mcal{V}$,
$(\psi, \varphi)_\mcal{V} = (\varphi, \psi)_\mcal{V}^*$, 
$\|\psi\|_\mcal{V}^2 \equiv (\psi, \psi)_\mcal{V} > 0$ for 
$\psi \neq 0$ and $\|\psi\|_\mcal{V} = 0$ for $\psi = 0$), we leave 
this scalar product arbitrary at this point. The scalar product turns 
$\mcal{V}$ into a separable Euclidean space, and we can use standard 
theorems \cite{Prugovecki} to complete this space, thereby defining a 
{\em separable Hilbert space\/} $\mcal{H}$. As a result, we have 
$\mcal{V} \subset \mcal{H}$ and
\beq
(\psi, \psi')_\mcal{H} = (\psi, \psi')_\mcal{V} \quad \mbox{for all } 
\psi, \psi' \in \mcal{V} \quad.
\eeq
(Recall that in addition to the finite superpositions of the $\psi_n$, 
$n = 0, \ldots$, that make up $\mcal{V}$, the Hilbert space $\mcal{H}$ 
also contains all limits $\psi = \lim_{n \to \infty} f_n$, 
$f_n \in \mcal{V}$, of Cauchy sequences of vectors of $\mcal{V}$.)

The Hilbert space $\mcal{H}$ is the natural choice as the space of
states for the system described by the Hamiltonian $H$, because it 
allows a consistent probabilistic interpretation of the model in terms
of scalar products of states.

It is well-known \cite{Prugovecki} that all infinite-dimensional,
separable Hilbert spaces are unitarily equivalent to the Hilbert space
$L_2(- \infty, \infty)$. This unitary equivalence means that an
isomorphism $T: \mcal{H} \to L_2$ exists that respects the scalar
products in both spaces:
\beq\label{2j}
(\psi, \psi')_\mcal{H} = (T \psi, T \psi')_{L_2} \quad \mbox{for all }
\psi, \psi' \in \mcal{H} \quad.
\eeq
The transformation $T$ is called a unitary transformation from 
$\mcal{H}$ onto $L_2$ \cite{Prugovecki}. (This notion has to be
distinguished from a unitary operator $U$ that is defined to be an 
automorphism, e.~g.\ $U: L_2 \to L_2$, that respects the norm, and 
therefore fulfills $U^\dagger = U^{- 1}$.) Given an operator 
$A_\mcal{H}$ that is defined in $\mcal{H}$, there is a corresponding 
operator 
\beq\label{2a}
A_{L_2} = T A_\mcal{H} T^{- 1}
\eeq
that is defined in $L_2$ and fulfills
\beq
(\psi, A_\mcal{H} \psi')_\mcal{H}
= (T \psi, A_{L_2} T \psi')_{L_2} 
\eeq
for all $\psi \in \mcal{H}$ and $\psi'$ in the domain of definition of 
$A_\mcal{H}$. If one denotes the Hermitian adjoint of an operator 
$A_\mcal{H}$ with respect to the scalar product in $\mcal{H}$ by 
$A_\mcal{H}^\ddagger$,
\beq
(\psi, A_\mcal{H} \psi')_\mcal{H} = (A_\mcal{H}^\ddagger \psi,
\psi')_\mcal{H} \quad,
\eeq
one finds 
\beq\label{2c}
A_{L_2}^\dagger = T A_\mcal{H}^\ddagger T^{- 1} \quad,
\eeq
so that, for example, $A_{L_2}$ is $L_2$ Hermitian if $A_\mcal{H}$ is 
$\mcal{H}$ Hermitian, and vice versa.

The equations (\ref{2a}) -- (\ref{2c}) imply that theories defined in 
$\mcal{H}$ and $L_2$ are physically indistinguishable.

\subsection{Hilbert spaces with Hermitian Hamiltonian}
\label{s21}

We have, up to this point, not specified the details of the scalar
product in $\mcal{H}$; the construction outlined above works with
every scalar product that can be defined on the space $\mcal{V}$. But
obviously not only the value of matrix elements, but the whole
structure of the space $\mcal{H}$ depends on this choice. And although
any choice allows a consistent probabilistic interpretation, we will 
here consider only Hilbert spaces in which {\em the Hamiltonian is 
Hermitian}. The reason is that we consider the system described by 
(\ref{2i}) as a closed system. (This appears to be an implicit 
assumption in most of the models treated recently.) According to a 
theorem by Wigner \cite{Wigner} the unitarity of the time evolution 
for such systems is a consequence of fundamental properties of quantum 
theories as, for example, the linearity of the time-evolution operator 
and the fact that physical states are described by rays 
$\{\lambda \psi; \lambda \in {\Bbb C}, \psi \in \mcal{H}\}$ rather 
than vectors $\psi$. If we want to keep these properties, and if the 
Hamiltonian $H$ is to be interpreted as the generator of time 
evolution, $H$ has to be Hermitian in the underlying Hilbert space
$\mcal{H}$.
 
In our context, it is the reality of the spectrum that makes the 
construction of Hilbert spaces with Hermitian $H$ possible. Define, 
for example,
\beq\label{2b}
(\psi_n, \psi_m)_\mcal{H} = \delta_{n m} \quad.
\eeq
It is easy to verify that (\ref{2b}) is a well-defined scalar product
in $\mcal{H}$. (Here the linear independence of the eigenstates
$\psi_n$ is important.) We have (from now on we denote $H$ by 
$H_\mcal{H}$)
\beq
(\psi_n, H_\mcal{H} \psi_m)_\mcal{H} = E_n \delta_{n m}
= (H_\mcal{H} \psi_n, \psi_m)_\mcal{H} \quad,
\eeq
and since the $\psi_n$ span the whole space, we can conclude
\beq\label{2o}
H_\mcal{H}^\ddagger = H_\mcal{H} \quad \mbox{and} \quad
H_{L_2}^\dagger = H_{L_2} \quad,
\eeq
where we have used (\ref{2c}). Thus, the choice (\ref{2b}) leads to 
Hermitian Hamiltonians $H_\mcal{H}$ and $H_{L_2}$ in their respective 
spaces.

Note that there is no problem to explicitly construct a transformation
$T: \mcal{H} \to L_2$ and the Hermitian operator $H_{L_2}$. It is
enough to take an arbitrary complete, orthonormalized set of $L_2$
functions, e.~g. the eigenstates $\varphi_n(x)$ of the harmonic 
oscillator, and define $T$ in (\ref{2j}) as the linear transformation 
that fulfills $T \psi_n = \varphi_n$. Then 
$H_{L_2} = T H_\mcal{H}(x, p) T^{- 1}$ has the required properties. 
The crucial point is that the transformed operators $x_{L_2}$ and 
$p_{L_2}$ will have the same complicated properties concerning their
Hermitian adjoints as $x$ and $p$ and will therefore in general be 
void of any physical significance. One has to find a transformation 
$T$ that leads to a Hamiltonian $H_{L_2} = H(x_{L_2}, p_{L_2})$ that 
allows a clear physical interpretation.

For achieving this, one may employ the uniqueness theorem by von 
Neumann \cite{Prugovecki}. It states that all irreducible 
representations of two self-adjoint operators $x^{\mrm{c}}$, 
$p^{\mrm{c}}$ that are defined in a separable Hilbert space and 
fulfill canonical commutation relations are unitarily equivalent. This 
means that given two such operators in a separable Hilbert space 
$\mcal{H}$, a unitary transformation $T: \mcal{H} \to L_2$ exists such 
that $x^{\mrm{c}}_{L_2} = T x^{\mrm{c}}_\mcal{H} T^{- 1}$ and 
$p^{\mrm{c}}_{L_2}$ are the two canonical $L_2$ operators
\beq
(x^{\mrm{c}}_{L_2} \varphi)(x) = x \varphi(x) \quad,\quad
(p^{\mrm{c}}_{L_2} \varphi)(x) = - \imu {\md \varphi \over \md x} 
\quad.
\eeq
In this way the uniqueness theorem allows one to find a unique (modulo
unitary equivalence) interpretation for quantum theories that are
defined on separable Hilbert spaces other than $L_2$.

Thus our aim is to find a set of canonical operators
$(x^{\mrm{c}}_\mcal{H}, p^{\mrm{c}}_\mcal{H})$ that are self-adjoint 
in $\mcal{H}$,
\beq
(x^{\mrm{c}}_\mcal{H})^\ddagger = x^{\mrm{c}}_\mcal{H} \quad,\quad
(p^{\mrm{c}}_\mcal{H})^\ddagger = p^{\mrm{c}}_\mcal{H} \quad,
\eeq
and fulfill
\beq
[x^{\mrm{c}}_\mcal{H}, p^{\mrm{c}}_\mcal{H}] = \imu \quad.
\eeq
If we are able to express $H_\mcal{H}$ as a function of these two
operators, 
$H_\mcal{H} = H(x^{\mrm{c}}_\mcal{H}, p^{\mrm{c}}_\mcal{H})$, we 
can immediately transform the model into the space $L_2$ with 
$H_{L_2} = H(x^{\mrm{c}}_{L_2}, p^{\mrm{c}}_{L_2})$. Here the physical 
meaning is clear, because one can interpret $x^{\mrm{c}}_{L_2}$ as the 
position-space observable and $p^{\mrm{c}}_{L_2}$ as the momentum 
(generator of $x^{\mrm{c}}_{L_2}$ translations).

We emphasize that this canonical formulation 
$H_{L_2} = H(x^{\mrm{c}}_{L_2}, p^{\mrm{c}}_{L_2})$ is not just an
alternative formulation for the original problem. In our opinion it is
the only formulation which has a clear physical meaning. Thus, if such
a formulation turns out to be impossible, we argue that the original
problem (\ref{2i}) is inconsistent. On the other hand, if a canonical
formulation is possible, then the non-Hermiticity of (\ref{2i}) with 
respect to the space $L_2$ is only a superficial one, due to a choice 
of variables that is inadequate for the given physical problem.

A necessary condition for the self-adjointness of the operators
$x^{\mrm{c}}$ and $p^{\mrm{c}}$ is their Hermiticity. In physical
considerations both notions are usually identified. We shall do the
same here, except in cases in which the difference becomes essential 
for the physical conclusions. These cases will be clearly indicated.

Of course, definition (\ref{2b}) is not the only scalar product giving
a Hermitian $H$. The condition that a scalar product $(.,.)'_\mcal{V}$ 
in the space $\mcal{V}$ leads to a Hermitian Hamiltonian $H$ follows 
from the requirement 
$(\psi_n, H \psi_m)'_\mcal{V} = (H \psi_n, \psi_m)'_\mcal{V}$ for all 
$n, m$, which gives $(E_n - E_m) (\psi_n,\psi_m)'_\mcal{V} = 0$. Thus
all scalar products for which the $\psi_n$ are mutually orthogonal 
lead to a Hermitian $H$. Therefore the most general ansatz for a 
scalar product with Hermitian $H$ is
\beq\label{2q}
(\psi_n,\psi_m)'_\mcal{V} = \gamma_n \delta_{n m} \quad,
\eeq
where the $\gamma_n$ are some positive constants. With respect to this 
scalar product, vectors $\hat{\psi}_n = \psi_n / c_n$ with 
$|c_n|^2 = \gamma_n$ are orthonormal. Let's define the Euclidean space
$\mcal{E}$ to be the space $\mcal{V}$ equipped with the scalar product
(\ref{2b}), and let $\hat{\mcal{E}}$ be the Euclidean space build from
$\mcal{V}$ and (\ref{2q}). Eventually, $\mcal{E}$ and $\hat{\mcal{E}}$
would be completed to give the Hilbert spaces $\mcal{H}$ and 
$\hat{\mcal{H}}$, resp. Every linear operator 
\beq
A: \mcal{V} \to \mcal{V} \quad,\quad A \psi_n 
= \sum_{n, m} a_{n m} \psi_m \quad,
\eeq
has a meaning in both spaces $\mcal{E}$, $\hat{\mcal{E}}$, but their
properties, for example concerning their Hermitian adjoints, will in
general be different in $\mcal{E}$ and $\hat{\mcal{E}}$. On the other 
hand, the Euclidean spaces $\mcal{E}$ and $\hat{\mcal{E}}$ are 
connected by the unitary transformation
$\hat{T}: \mcal{E} \to \hat{\mcal{E}}$,
\beq\label{2r}
\hat{T} \psi_n = \hat{\psi}_n = {\psi_n \over c_n} \quad,\quad
(\psi_n, \psi_m)_\mcal{V} 
= (\hat{T} \psi_n, \hat{T} \psi_m)'_\mcal{V} \quad,
\eeq
and therefore $A$ as an operator $\mcal{E} \to \mcal{E}$ can be
mapped to
\beq
\hat{A} = \hat{T} A \hat{T}^{- 1}: \hat{\mcal{E}} \to \hat{\mcal{E}} 
\quad.
\eeq
These operators, $A: \mcal{E} \to \mcal{E}$ and
$\hat{A}: \hat{\mcal{E}} \to \hat{\mcal{E}}$, share the same 
properties. The transformation $\hat{T}$ in (\ref{2r}) is only a 
function of $H$, so that $\hat{H} = H$. But in general, unless $A$ 
commutes with $\hat{T}$, one has $\hat{A} \neq A$.

This means that to a certain extent one can adjust the properties of 
an operator $A: \mcal{V} \to \mcal{V}$ by choosing an appropriate 
scalar product. This will turn out to be helpful for the construction
of sets of canonical operators (Section~\ref{s32}). If, for example, 
an operator $F(H)^{- 1} A F(H): \mcal{E} \to \mcal{E}$ is Hermitian in 
$\mcal{E}$, then $A$ is Hermitian in the space $\hat{\mcal{E}}$ which 
is defined by $\hat{\psi}_n = F(H) \psi_n$ and
$(\hat{\psi}_n, \hat{\psi}_m)'_\mcal{V} = \delta_{n m}$.

\subsection{Scalar products along complex paths}
\label{s22}

In the work of Bender and collaborators \cite{Bender...} the 
eigenfunctions $\psi_n(z)$ are defined with $z$ describing a curve in 
the complex plane which is chosen in such a way that the 
eigenfunctions are decreasing asymptotically. Let a parameterization 
for this curve be $z = z(s)$, $- \infty < s < \infty$. We may then
define a scalar product by
\beq\label{2k}
(\psi, \psi')_\mcal{H} = \int \limits_{- \infty}^\infty \md s \,
[\psi(z(s))]^* \psi'(z(s)) \quad.
\eeq
The question is whether such scalar products can lead to a Hermitian 
Hamiltonian.

According to our general discussion, $\psi(z)$ is an element of 
$\mcal{H}$ and $\varphi(s) = \psi(z(s)) \in L_2$, i.~e.\ the
transformation $T: \mcal{H} \to L_2$ acts as 
\beq\label{2p}
(T \psi)(s) = \psi(z(s))= \varphi(s)
\eeq
and
\beq
(\psi, \psi')_\mcal{H} = (T \psi, T \psi')_{L_2} \quad.
\eeq

Consider two sets of operators: $x_\mcal{H}$ and $p_\mcal{H}$ act in
$\mcal{H}$ in the usual way:
\beq
(x_\mcal{H} \psi)(z) = z \psi(z) \quad,\quad
(p_\mcal{H} \psi)(z) = - \imu {\md \psi \over \md z}
\eeq
and can be transformed into $x_{L_2}$ and $p_{L_2}$ which act in a 
complicated way. The second set consists of the canonical operators
$x^{\mrm{c}}_{L_2}$ and $p^{\mrm{c}}_{L_2}$ that act in $L_2$ like
\beq
(x^{\mrm{c}}_{L_2} \varphi)(s) = s \varphi(s) \quad,\quad
(p^{\mrm{c}}_{L_2} \varphi)(s) = - \imu {\md \varphi \over \md s} 
\quad,
\eeq
are Hermitian and can be transformed into $x^{\mrm{c}}_\mcal{H}$ and
$p^{\mrm{c}}_\mcal{H}$. With these operators, one finds
\begin{eqnarray}
(\psi, x_\mcal{H} \psi')_\mcal{H} 
& = & \int \md s \, \varphi^*(s) z(s) \varphi'(s) \\
& = & \int \md s \, \varphi^*(s) (z(x^{\mrm{c}}_{L_2}) \varphi')(s) \\
& = & (\varphi, z(x^{\mrm{c}}_{L_2}) \varphi')_{L_2} 
\equiv (\varphi, x_{L_2} \varphi')_{L_2} \quad,
\end{eqnarray}
i.~e.\ $x_{L_2} = T x_\mcal{H} T^{- 1} = z(x^{\mrm{c}}_{L_2})$, which 
is the path $z(s)$ with the real parameter $s$ replaced by the 
Hermitian operator $x^{\mrm{c}}_{L_2}$. This is of course also true 
for the operators in $\mcal{H}$:
\beq\label{2l}
x_\mcal{H} = z(x^{\mrm{c}}_\mcal{H}) \quad.
\eeq
The operator $p_\mcal{H}$ can be analysed analogously: Using 
$\md \psi / \md z = (\md \varphi / \md s) / (\md z / \md s)$, one 
obtains
\begin{eqnarray}
\lefteqn{(\psi, p_\mcal{H} \psi')_\mcal{H} 
= - \imu \int \md s \, [\psi(z(s))]^* 
\left. {\md \psi' \over \md z} \right|_{z = z(s)}} \nonumber \\
& = & \int \md s \, \varphi^*(s) 
\left( \! {1 \over \md z(x^{\mrm{c}}_{L_2}) / \md x^{\mrm{c}}_{L_2}} 
p^{\mrm{c}}_{L_2} \varphi' \! \right)\!(s) \nonumber \\
& = & (\varphi, {1 \over \md z / \md x^{\mrm{c}}_{L_2}} 
p^{\mrm{c}}_{L_2} \varphi')_{L_2} 
\equiv (\varphi, p_{L_2} \varphi')_{L_2} \quad,
\end{eqnarray}
and consequently
\beq\label{2m}
p_\mcal{H} 
= {1 \over \md z(x^{\mrm{c}}_\mcal{H}) / \md x^{\mrm{c}}_\mcal{H}} 
p^{\mrm{c}}_\mcal{H} \quad.
\eeq
Hence definition (\ref{2k}) leads to simple relations connecting the
original operators $(x_\mcal{H}, p_\mcal{H})$ with the canonical ones
$(x^{\mrm{c}}_\mcal{H}, p^{\mrm{c}}_\mcal{H})$. (But note that 
(\ref{2k}) explicitly depends on the parameterization chosen. This
leaves a certain freedom for the definition of the canonical 
operators.)

The Hermiticity of the canonical operators immediately leads to
\beq
x_\mcal{H}^\ddagger = z^*(x^{\mrm{c}}_\mcal{H}) 
\eeq
and with 
$\md x_\mcal{H} / \md x^{\mrm{c}}_\mcal{H} p_\mcal{H} 
= p_\mcal{H}^\ddagger \md x_\mcal{H}^\ddagger 
/ \md x^{\mrm{c}}_\mcal{H}$ 
and 
$[p_\mcal{H}, f(x_\mcal{H})] = - \imu \md f / \md x_\mcal{H}$ one
obtains
\beq
p_\mcal{H}^\ddagger 
= v(x^{\mrm{c}}_\mcal{H}) [\imu w(x^{\mrm{c}}_\mcal{H}) + p_\mcal{H}]
\eeq
with
\begin{eqnarray}
v(x^{\mrm{c}}_\mcal{H}) 
& = & {\md x_\mcal{H} / \md x^{\mrm{c}}_\mcal{H}
\over \md x_\mcal{H}^\ddagger / \md x^{\mrm{c}}_\mcal{H}} \quad,\quad
\label{2d} \\
w(x^{\mrm{c}}_\mcal{H}) 
& = & {1 \over (\md x_\mcal{H} / \md x^{\mrm{c}}_\mcal{H}) 
(\md x_\mcal{H}^\ddagger / \md x^{\mrm{c}}_\mcal{H})} 
{\md^2 x_\mcal{H}^\ddagger \over (\md x^{\mrm{c}}_\mcal{H})^2} \quad.
\label{2e}
\end{eqnarray}
One can check that $(p_\mcal{H}^\ddagger)^\ddagger = p_\mcal{H}$; this 
gives the unitarity of $v$:
\beq
v(x^{\mrm{c}})^\ddagger = v(x^{\mrm{c}})^{- 1}
\eeq
and an additional condition:
\beq\label{2f}
{\md v \over \md x^{\mrm{c}}_\mcal{H}} 
= \left[ w(x^{\mrm{c}}_\mcal{H})^\ddagger - v(x^{\mrm{c}}_\mcal{H}) 
w(x^{\mrm{c}}_\mcal{H}) \right] {\md x_\mcal{H} 
\over \md x_\mcal{H}^{\mrm{c}}} \quad.
\eeq
These equations are identically fulfilled for (\ref{2d}) and 
(\ref{2e}).

Now one can ask whether a curve or parameterization exists that makes 
a given Hamiltonian Hermitian. Consider a Hamiltonian of the form
\beq\label{2n}
H(x_\mcal{H}, p_\mcal{H}) = p_\mcal{H}^2 + V(x_\mcal{H}) \quad.
\eeq
In its Hermitian conjugate 
$H^\ddagger = (p_\mcal{H}^\ddagger)^2 + (V(x_\mcal{H}))^\ddagger$ the
operator $(p_\mcal{H}^\ddagger)^2$ is given by
\begin{eqnarray}
(p_\mcal{H}^\ddagger)^2 
& = & \left[ v \left( {\md \over \md x_\mcal{H}} v w \right) 
- (v w)^2 \right] \label{2g} \\
& & {} + \imu \left[ 2 v^2 w - v {\md v \over \md x_\mcal{H}} \right] 
p_\mcal{H} + v^2 p_\mcal{H}^2 \quad. \nonumber
\end{eqnarray}
Since $(V(x_\mcal{H}))^\ddagger$ does not contain $p_\mcal{H}$, a 
necessary condition for the Hermiticity of $H$ is that the term linear 
in $p_\mcal{H}$ in (\ref{2g}) vanishes. This, together with relation 
(\ref{2f}) gives $3 v w = w^\ddagger$. Taking the Hermitian adjoint of 
this equation and using the unitarity of $v$ then shows that 
necessarily $w = 0$. Since $w$ contains 
$\md^2 x_\mcal{H}^\ddagger / (\md x^{\mrm{c}}_\mcal{H})^2$, which is 
the operator version of $\md^2 z^* / \md s^2$, one sees that only 
straight lines,
\beq\label{2h}
z(s) = a + b s \quad,\quad a, b \in {\Bbb C} \quad,
\eeq
are allowed. Then $v$ is a constant phase, 
$v(x_\mcal{H}) = b / b^* = \e^{2 \imu \arg b}$ and 
$(p_\mcal{H}^\ddagger)^2 = \e^{4 \imu \arg b} p_\mcal{H}^2$. Hence
$p_\mcal{H}^2$ is only Hermitian if the parameter $b$ satisfies
\beq
\arg b = n {\pi \over 2} \quad.
\eeq
Under this condition, one has
\beq
H^\ddagger = p_\mcal{H}^2 + (V(x_\mcal{H}))^\ddagger \quad,
\eeq
and it may happen that $a$ and $|b|$ can be chosen in such a way that
also the potential becomes Hermitian. But the fact that only straight
lines in the complex plane can lead to Hermitian Hamiltonians of the
form (\ref{2n}) seems to restrict the usefulness of scalar products 
along complex paths severely.

In the simple case of straight lines one can explicitly construct the
transformation $T: \mcal{H} \to L_2$ (cf.\ (\ref{2p})): it is given 
by a combination
\beq
T = {1 \over \sqrt{b}} T_{\mrm{t}}(a / b) T_{\mrm{d}}(\ln b)
\eeq
of a (complex) translation
\beq
T_{\mrm{t}}(a / b) = \exp \imu (a / b) p_\mcal{H} \,,\quad
(T_{\mrm{t}}(a / b) \psi)(s) = \psi(s + a / b) \,,
\eeq
and a (complex) dilatation
\begin{eqnarray}
T_{\mrm{d}}(\ln b) 
& = & \exp \left( {\imu \over 2} (x_\mcal{H} p_\mcal{H} 
+ p_\mcal{H} x_\mcal{H}) \ln b \right) \,, \\
(T_{\mrm{d}}(\ln b) \psi)(s) & = & \sqrt{b} \, \psi(b s) \quad.
\end{eqnarray}

\section{Discussion of various models}
\label{s3}

Unfortunately, most works dealing with non-Hermitian Hamiltonians are
based on numerical methods which are not very well adapted for studies
of the structure of the space $\mcal{H}$. But note that an 
investigation in this direction has been started in \cite{Bender2}. 
We will here consider exactly solvable models. Among those works that 
also treat exactly solvable models, we mention \cite{Beckers}, where a 
similar construction of a Hilbert space as the one outlined in 
Section~\ref{s2} has been recently applied to the study of certain 
coherent states.

In the following we will throughout suppress the subscript $\mcal{H}$
for operators that act in the space $\mcal{H}$; for operators acting 
in $L_2$ we keep the subscript $L_2$.

\subsection{The one-dimensional Coulomb problem on the real half axis} 
\label{s31} 

Let us start with a very simple example. We first discuss it in a way 
inspired by Bender et al.\ \cite{Bender...,Bender3} and then 
reinterpret it according to the framework of Section~\ref{s2}. 

Consider the one-dimensional Coulomb problem on the real half axis 
$x > 0$: The Hamiltonian reads
\beq
H^{(0)}_{L_2}(a) 
= (p^{\mrm{c}}_{L_2})^2 - {a \over x^{\mrm{c}}_{L_2}} \quad,
\eeq
the eigenfunctions for real $x > 0$ and $a > 0$ that are finite at the 
origin, have a finite slope there, and are decreasing for 
$x \to \infty$ are \cite{Loudon}
\beq\label{3s}
\Phi_n(x; a) 
= \e^{- a x / (2 n)} {a x \over n} L^{(1)}_{n - 1}( a x / n) 
\quad,
\eeq
$n = 1, \ldots$, with the Laguerre polynomials $L^{(1)}_{n - 1}$. The 
eigenvalues are
\beq
E_n^{(0)}(a) = - \left( {a \over 2 n} \right)^2 \quad.
\eeq
It is clear that $\Phi_n(x; a)$ is normalizable in $L_2(0, \infty)$ 
for real and positive $a$; the spectrum in this case is a bound-state 
spectrum. 

We may now use the solutions (\ref{3s}) with $a = \imu \alpha$,
$\alpha$ real and positive,
\beq\label{3m}
\psi_n(x)
= \Phi_n(x; \imu \alpha)
= \e^{- \imu \alpha x / (2 n)} {\imu \alpha x \over n}
L^{(1)}_{n - 1}(\imu \alpha x / n) \quad.
\eeq
The energies are then still real,
\beq\label{3k}
E_n = \left( {\alpha \over 2 n} \right)^2 > 0 \quad,
\eeq
and the Hamiltonian is
\beq\label{3b}
H = p^2 - {\imu \alpha \over x} \quad,
\eeq
thus non-Hermitian with respect to $L_2$. The asymptotic behavior of 
$\psi_n(x)$ for complex $x = |x| \e^{\imu \vartheta}$ and large $|x|$ 
is given by
\beq
|\psi_n(x)| \sim \exp \left( {\alpha |x| \over 2 n} \sin \vartheta
\right) \quad,
\eeq
hence $\psi_n$ converges most rapidly on the anti-Stokes line given 
by $\vartheta = 3 \pi / 2$. The eigenfunctions (\ref{3m}) have no 
finite norm in $L_2(0, \infty)$. One would therefore consider the 
$\psi_n(x)$ on the anti-Stokes line
\beq
x(s) = \e^{\imu \vartheta} s \quad,\quad 0 \leq s < \infty \quad. 
\eeq

In terms of our general discussion in Section~\ref{s22}, we would 
define the scalar product in the space $\mcal{H}$ by (cf.~(\ref{2k}))
\beq
(\psi, \psi')_\mcal{H} 
= \int \limits_0^\infty \md s \, [\psi(\e^{\imu \vartheta} s)]^*
\psi'(\e^{\imu \vartheta} s) \quad.
\eeq
This corresponds to the special case (\ref{2h}) with $a = 0$ and
$b = \e^{\imu \vartheta}$. The canonical operators are given by 
(\ref{2l}) and (\ref{2m}):
\beq
x^{\mrm{c}} = \e^{- \imu \vartheta} x \quad,\quad
p^{\mrm{c}} = \e^{\imu \vartheta} p \quad,
\eeq
therefore the canonical form of the Hamiltonian (\ref{3b}) is
\beq\label{3j}
H(x^{\mrm{c}}, p^{\mrm{c}}) 
= \e^{- 2 \imu \vartheta} (p^{\mrm{c}})^2 
- \imu \e^{- \imu \vartheta} {\alpha \over x^{\mrm{c}}}
= - (p^{\mrm{c}})^2 + {\alpha \over x^{\mrm{c}}}
\eeq
and is obviously Hermitian. Its $L_2$ realization via 
$T: \mcal{H} \to L_2$, 
$\varphi(s) = (T \psi)(s) = \psi(\e^{\imu \vartheta} s)$, is
\beq\label{3l}
H(x^{\mrm{c}}_{L_2}, p^{\mrm{c}}_{L_2})
= - (p^{\mrm{c}}_{L_2})^2 + {\alpha \over x^{\mrm{c}}_{L_2}} 
= - H^{(0)}_{L_2}(\alpha) \quad,
\eeq
and the transformed eigenfunctions read
\beq
\varphi_n(s) = \Phi_n(s; \alpha) \quad.
\eeq

In summary, the quantum-mechanical interpretation based on the 
canonical form (\ref{3j}) is completely equivalent to the original, 
Hermitian problem; the change of the sign of the spectrum (\ref{3k}) 
is reflected by the sign in (\ref{3l}). Hence the solutions (\ref{3m}) 
do not contain any new information. Moreover, the kinetic energy in
(\ref{3j}) has---contrary to (\ref{3b})---the wrong sign. This makes 
the model unphysical.

\subsection{The model of Cannata et al.}
\label{s32}

In the model introduced by Cannata, Junker and Trost \cite{Cannata}
the Hamiltonian is
\beq\label{3f}
H = {1 \over 2} (p^2 + \e^{2 \imu x}) 
\eeq
(a similar model has been considered in \cite{Bender4}). The 
Schr\"o\-dinger equation for this Hamiltonian can be reduced to 
Bessel's differential equation, so that its general solution is given 
by
\beq
\Psi_\nu(x) = c_1 H^{(1)}_\nu(\e^{\imu x}) 
+ c_2 H^{(2)}_\nu(\e^{\imu x}) \quad,\quad
E_\nu = {\nu^2 \over 2} \quad.
\eeq
For real $x$ no normalizable solutions exist. But if $x$ is chosen
along curves in the complex plane, different possibilities occur. We
briefly summarize the main cases; details can be found in
\cite{Cannata}.

In the upper-half $x$ plane, one can find normalizable solutions for
every real value of the energy, i.~e.\ no quantization condition
appears and the spectrum has no lower bound so that the system is
unphysical. If one divides the lower-half $x$ plane into vertical 
strips separated by the lines $\re x = n \pi$ with integer $n$, and if 
one assumes that the curves along which the wave functions are defined 
have two asymptotes that are vertical lines going down, then the 
following configurations occur: If a curve starts and ends in the same 
strip, no quantization condition appears, the spectrum is again 
unbounded from below. If the curve starts in one strip and ends in the 
next strip, then no normalizable solutions exist. If, however, the 
curve starts in one strip and ends in the next-to-next strip (e.~g. 
one asymptote of the curve is the anti-Stokes line at 
$\re x = \pi / 2$ and the other asymptote is at $\re x = 5 \pi / 2$), 
the quantization condition $\nu = n + 1 / 2$, $n$ integer, appears. 
One then has the solution
\beq\label{3n}
\psi_n(x) = H^{(1)}_{n + 1 / 2}(\e^{\imu x}) \quad,\quad
E_n = {(n + 1 / 2)^2 \over 2} \quad.
\eeq
In addition to this case, on which our subsequent discussion 
concentrates, there exist other curves with similar spectra 
\cite{Cannata}.

It follows from (\ref{3n}) that 
\beq\label{3al}
\psi_n = \imu (- 1)^{- n - 1} \psi_{- n - 1} \quad.
\eeq
This means that the $\psi_n$ with $n \geq 0$ constitute a set of 
linearly independent eigenstates, so that one can start from the space 
$\mcal{V} = \vspan \{\psi_0,\psi_1, \ldots\}$ to construct $\mcal{H}$.

The recursion relations 
$z \md Z_\nu / \md z \pm \nu Z_\nu = \pm z Z_{\nu \mp 1}$ for Bessel 
functions can be used to find
\begin{eqnarray}
\e^{- \imu x} \psi_n 
& = & {1 \over 2 n + 1} (\psi_{n - 1} + \psi_{n + 1}) \quad, 
\label{3c} \\
\e^{- \imu x} p \psi_n
& = & {1 \over 2} ( \psi_{n - 1} - \psi_{n + 1}) \quad. \label{3d}
\end{eqnarray}
Thus the two operators $\e^{- \imu x}$ and $\e^{- \imu x} p$ are 
well-defined operators $\mcal{H} \to \mcal{H}$. The two relations 
(\ref{3c}) and (\ref{3d}) lead to the version
\beq
\e^{- 2 \imu x} \left[ (n + 1 / 2)^2 - p^2 \right] \psi_n = \psi_n
\eeq
of the Schr\"odinger equation $H \psi_n = E_n \psi_n$. (Note that 
these are algebraic relations that hold for all Bessel functions and 
all $n \in \mathbb{C}$. It is (\ref{3al}) and the definition of 
$\mcal{V}$ that are specific to the solution (\ref{3n}).)

We first consider the question whether one can define a scalar product
along a curve such that the Hamiltonian (\ref{3f}) becomes Hermitian.
Since (\ref{3f}) is of the form (\ref{2n}), it is clear that only
straight lines $x = a + b x^{\mrm{c}}$ are possible. Then 
$p^{\mrm{c}} = b p$, and the Hamiltonian can be written as 
\beq
H(x^{\mrm{c}}, p^{\mrm{c}})
= {(p^{\mrm{c}})^2 \over 2 b^2} 
+ {\e^{2 \imu (a + b x^{\mrm{c}})} \over 2} \quad.
\eeq
If one wants this to be Hermitian, only lines with $\re a = n \pi / 2$
and $\re b = 0$ are allowed. Although such lines may be the asymptotes
of curves that lead to the solution (\ref{3n}), one cannot get a
connected curve by joining these lines. 

Therefore, we now return to the general discussion of 
Section~\ref{s21} and define the Hilbert space $\mcal{H}$ by imposing 
the orthogonality condition 
$(\psi_n, \psi_m)_\mcal{H} = \delta_{n m}$ for $n, m \geq 0$, 
see (\ref{2b}). The equations (\ref{3c}) and (\ref{3d}) can then be 
used to calculate the adjoints of $\e^{- \imu x}$ and 
$\e^{- \imu x} p$; the results are
\beq
\e^{\imu x^\ddagger} \psi_n 
= {1 \over 2 (n - 1) + 1} \psi_{n - 1} + {1 \over 2 (n + 1) + 1} 
\psi_{n + 1}
\eeq
and
\beq
\left( \e^{- \imu x} p \right)^\ddagger = - \e^{- \imu x} p \quad.
\eeq
It follows that $\tilde{p}^{\mrm{c}} = \imu \e^{- \imu x} p$ is 
Hermitian.  Furthermore, $[x, p] = \imu$ leads to 
$[\e^{\imu x}, \tilde{p}^{\mrm{c}}] = - \imu$. Hence the two operators
\beq
\tilde{x}^{\mrm{c}} 
= - {1 \over 2} \left( \e^{\imu x} + \e^{- \imu x^\ddagger} \right) 
\quad,\quad 
\tilde{p}^{\mrm{c}} = \imu \e^{- \imu x} p
\eeq
are Hermitian and fulfill 
$[\tilde{x}^{\mrm{c}}, \tilde{p}^{\mrm{c}}] = \imu$ so that they form 
a canonical set. 

But it seems that one cannot write the Hamiltonian (\ref{3f}) only as 
a function 
of $\tilde{x}^{\mrm{c}}$ and $\tilde{p}^{\mrm{c}}$. One can readily 
express $H$ as 
$H = H(\tilde{x}^{\mrm{c}}, \tilde{p}^{\mrm{c}}, 
\tilde{x}^{\mrm{c} \prime})$, where 
$\tilde{x}^{\mrm{c} \prime} 
= \imu (\e^{\imu x} - \e^{- \imu x^\ddagger}) / 2$ 
is another Hermitian operator. This operator fulfills 
$[\tilde{x}^{\mrm{c} \prime}, \tilde{p}^{\mrm{c}}] = 0$. This may mean 
that $\tilde{x}^{\mrm{c} \prime}$ is only a function of 
$\tilde{p}^{\mrm{c}}$, but we were not able to show this.

One therefore has to find other canonical variables. We emphasize that
for the following discussion we use the term Hermitian in the precise
sense of {\em symmetry}, i.~e.\ an operator $A$ is called Hermitian if 
$(\psi, A \psi')_\mcal{H} = (A \psi, \psi')_\mcal{H}$ for all
$\psi, \psi'$ in the domain of definition of $A$.

We can use the method described at the end of Section~\ref{s21} to try
to find a new scalar product that makes $H$ and $\e^{- \imu x}$
simultaneously Hermitian. As such a new scalar product is connected to
the old one by 
$(\psi_n, \psi_m)_\mcal{H}' = \gamma_n (\psi_n, \psi_m)_\mcal{H}$, 
$\gamma_n > 0$ (see (\ref{2q})), a Hermitian $\e^{- \imu x}$ would 
require real matrix elements $(\psi_n, \e^{- \imu x} \psi_n)_\mcal{H}$ 
for all $n$. But since $\psi_{- 1} = \imu \psi_0$, one obtains 
$(\psi_0, \e^{- \imu x} \psi_0)_\mcal{H} = \imu$, so that one cannot 
construct a scalar product for which $\e^{- \imu x}$ is Hermitian in 
the entire space $\mcal{H}$. The best one can do is to make 
$\e^{- \imu x}$ Hermitian in the subspace spanned by
$\{\psi_1, \psi_2, \ldots\}$. This is achieved by defining
\beq
\hat{\psi}_n = \sqrt[4]{H} \, \psi_n 
= \sqrt{n + 1 / 2} \, \psi_n / \sqrt[4]{2} \quad \mbox{for } n \geq 0 
\quad,
\eeq
and introducing a new Hilbert space $\hat{\mcal{H}}$ with the new 
scalar product 
$(\hat{\psi}_n, \hat{\psi}_m)_{\hat{\mcal{H}}} = \delta_{nm}$
for $n, m \geq 0$. Then
\beq\label{3e}
\e^{- \imu x} \hat{\psi}_n
= {1 \over 2 \sqrt{n + 1 / 2}}
\left( \! {\hat{\psi}_{n - 1} \over \sqrt{n - 1 + 1 / 2}}
+ {\hat{\psi}_{n + 1} \over \sqrt{n + 1 + 1 / 2}} \! \right)
\eeq
for $n \ge 1$ and
\beq\label{3ap}
\e^{- \imu x} \hat{\psi}_0
= \imu \hat{\psi}_0 + {\hat{\psi}_1 \over \sqrt{3}} \quad.
\eeq
These equations show 
\beq\label{3ar}
(\hat{\psi}_n, \e^{- \imu x} \hat{\psi}_m)_{\hat{\mcal{H}}}
= (\e^{- \imu x} \hat{\psi}_n, \hat{\psi}_m)_{\hat{\mcal{H}}}
\eeq
for all $n, m \geq 0$ except for $n = m = 0$, i.~e.\ the restriction
\beq
\left. \e^{- \imu x} \right|_{\hat{\mcal{V}}_1}:
\hat{\mcal{V}}_1 \to \hat{\mcal{H}}
\eeq
of $\e^{- \imu x}$ to the space
$\hat{\mcal{V}}_1 = \vspan \{ \hat{\psi}_1, \ldots\}$ is Hermitian. If 
we define
\beq\label{3aq}
\xi_n 
= {\hat{\psi}_{n - 1} \over \sqrt{n - 1 + 1 / 2}}
+ {\hat{\psi}_{n + 1} \over \sqrt{n + 1 + 1 / 2}} \quad \mbox{for } 
n \ge 1 \quad,
\eeq
then the image of $\hat{\mcal{V}}_1$ under $\e^{- \imu x}$ is given by
$\hat{\mcal{V}}_\xi = \vspan \{\xi_1, \ldots\}$. Consider the series
\begin{eqnarray}
\lefteqn{f^{(n)}_N = \sum_{k = 0}^N (- 1)^k \xi_{n + 2 k + 1}} 
\label{3aa} \\
& = & {\hat{\psi}_n \over \sqrt{n + 1 / 2}}
+ (- 1)^N {\hat{\psi}_{n + 2 (N + 1)} 
\over \sqrt{[n + 2 (N + 1)] + 1 / 2}} \quad.
\end{eqnarray}
It fulfills
\begin{eqnarray}
\lefteqn{\left\| f^{(n)}_N - \hat{\psi}_n / \sqrt{n + 1 / 2}
\right\|_{\hat{\mcal{H}}}} \\
& = & {1 \over \sqrt{[n + 2 (N + 1)] + 1 / 2}} \to 0 \quad \mbox{for }
N \to \infty \nonumber
\end{eqnarray}
and therefore converges in $\hat{\mcal{H}}$, i.~e.
\beq\label{3r}
{\hat{\psi}_n \over \sqrt{n + 1 / 2}}
= \sum_{k = 0}^\infty (- 1)^k \xi_{n + 2 k + 1} \quad \mbox{for }
n \geq 0 \quad,
\eeq
which means that the linearly independent (although non-orthogonal)
vectors $\xi_n$, $n \ge 1$, lie dense in the space $\hat{\mcal{H}}$. 
We now define the operator
\beq\label{3o}
\hat{x}^{\mrm{c}} 
= \left( \left. \e^{- \imu x} \right|_{\hat{\mcal{V}}_1} 
\right)^{- 1}: \hat{\mcal{V}}_\xi \to \hat{\mcal{V}}_1 \quad.
\eeq
This operator acts as
\beq\label{3g}
\hat{x}^{\mrm{c}} \xi_n = 2 \sqrt{n + 1 / 2} \, \hat{\psi}_n \quad,
\eeq
has a dense domain of definition in $\hat{\mcal{H}}$ and is Hermitian:
\beq\label{3an}
(\xi_n, \hat{x}^{\mrm{c}} \xi_m)_{\hat{\mcal{H}}}
= 2 (\delta_{n, m + 1} + \delta_{n + 1, m})
= (\hat{x}^{\mrm{c}} \xi_n, \xi_m)_{\hat{\mcal{H}}}
\eeq
for all $n, m \geq 1$. We use $\hat{x}^{\mrm{c}}$ as one of our 
canonical variables. Its conjugate momentum can be defined by
\beq\label{3i}
\hat{p}^{\mrm{c}} 
= - {\imu \over 2} (\e^{- \imu x} p + p \, \e^{- \imu x}) \quad.
\eeq

The operator $\hat{x}^{\mrm{c}}$ has to be distinguished from the
operator $\e^{\imu x}$. Furthermore, although $\hat{x}^{\mrm{c}}$ is
well-defined in the space $\hat{\mcal{V}}_\xi$ by (\ref{3g}), in
acting on the eigenfunctions $\hat{\psi}_n$ via (\ref{3r}) it leads to 
the divergent series
\begin{eqnarray}
\hat{x}^{\mrm{c}} \hat{\psi}_n
& = & 2 \sqrt{n + 1 / 2} \label{3ao} \\
& & \times \sum_{k = 0}^ \infty (- 1)^k \sqrt{n + 2 k + 1 + 1 / 2} \,
\hat{\psi}_{n + 2 k + 1} \nonumber
\end{eqnarray}
so that
$\hat{x}^{\mrm{c}} \hat{\psi}_n$ cannot be defined by this series.
Because of this, the construction of (\ref{3i}) and the demonstration 
of the commutation relation 
$[\hat{x}^{\mrm{c}}, \hat{p}^{\mrm{c}}] = \imu$ is a rather technical 
problem. The corresponding proofs are described in detail in 
Appendix~\ref{sa}. Here we just note that
\beq\label{3h}
\hat{p}^{\mrm{c}} \hat{\psi}_n
= {\imu \over 2 \sqrt{n + 1 / 2}} \left( \!
{- n \hat{\psi}_{n - 1} \over \sqrt{n - 1 + 1 / 2}}
+ {(n + 1) \hat{\psi}_{n + 1} \over \sqrt{n + 1 + 1 / 2}} \! \right)
\eeq
for all $n \geq 0$, and that indeed
\beq
(\xi_n, \hat{p}^{\mrm{c}} \xi_m)_{\hat{\mcal{H}}} 
= (\hat{p}^{\mrm{c}} \xi_n, \xi_m)_{\hat{\mcal{H}}} \quad 
\mbox{for all } n, m \geq 1 
\eeq
and
\beq
[\hat{x}^{\mrm{c}}, \hat{p}^{\mrm{c}}] \xi_n = \imu \xi_n \quad
\mbox{for all } n \geq 1 \quad.
\eeq

Now we are finally in the position to write down the canonical form of
the Hamiltonian of the model of Cannata et~al.: From (\ref{af}) we
have $p = \imu \hat{x}^{\mrm{c}} \hat{p}^{\mrm{c}} + 1 / 2$, and as
explained in Appendix~\ref{sa}, one can replace $\e^{\imu x}$ by 
$\hat{x}^{\mrm{c}}$ in the Hamiltonian (\ref{3f}), so that $H$ can be 
written as
\beq\label{3am}
H = H(\hat{x}^{\mrm{c}}, \hat{p}^{\mrm{c}})
= {1 \over 2} \left[ - (\hat{x}^{\mrm{c}})^2 (\hat{p}^{\mrm{c}})^2
+ 2 \imu \hat{x}^{\mrm{c}} \hat{p}^{\mrm{c}} + {1 \over 4}
+ (\hat{x}^{\mrm{c}})^2 \right] \,.
\eeq

One thing should be noted here: Although (\ref{3f}) multiplied with
$\hat{\psi}_n$ is of course well-defined in $\hat{\mcal{H}}$, the two
terms $p^2 \hat{\psi}_n$ and $\e^{2 \imu x} \hat{\psi}_n$ are
divergent; their divergencies cancel in the sum. With respect to
(\ref{3am}) this means that one can group the terms on the right-hand
side in such a way that no divergencies appear in intermediate 
results. This is achieved by writing (see (\ref{ad}))
\beq
H = {1 \over 2} \left[ \hat{x}^{\mrm{c}} \left[ 
\hat{x}^{\mrm{c}} \left( 1 - (\hat{p}^{\mrm{c}})^2 \right) 
+ 2 \imu \hat{p}^{\mrm{c}} \right] + {1 \over 4} \right] \quad.
\eeq
This form of the Hamiltonian is an operator that maps 
$\hat{\mcal{V}} = \vspan \{\hat{\psi}_0, \hat{\psi}_1, \ldots\}$ 
to itself. In this sense, (\ref{3am}) is equivalent to (\ref{3f}) in 
the entire space $\hat{\mcal{V}}$.

Let's consider the realization of this Hamiltonian in 
$L_2(- \infty, \infty)$, i.~e.\ 
$H_{L_2} = H(x^{\mrm{c}}_{L_2}, p^{\mrm{c}}_{L_2})$, in which 
$x^{\mrm{c}}_{L_2}$ and $p^{\mrm{c}}_{L_2}$ are identified with the 
canonical $L_2$ operators: 
\beq\label{3p}
H_{L_2}
= {1 \over 2} \left[ x^2 {\md^2 \over \md x^2} + 2 x {\md \over \md x}
+ {1\over 4} + x^2 \right] \quad.
\eeq
The general solution of the Schr\"odinger equation 
$H_{L_2} \Phi_\nu \allowbreak = (\nu + 1 / 2)^2 \Phi_\nu / 2$ is given 
by \cite{Gradstein} 
$\Phi_\nu(x) = c_1 \varphi_\nu(x) + c_2 \varphi_{- \nu - 1}(x)$ with
\beq\label{3ak}
\varphi_\nu(x) = {1 \over \sqrt{x}} J_{\nu + 1 / 2}(x) \quad,\quad 
\nu \in {\Bbb C} \quad.
\eeq
For the case corresponding to the spectrum (\ref{3n}), namely
integer $\nu = n$, these solutions are proportional to spherical 
Bessel functions $c_1 j_n(x) + c_2 j_{- n - 1}(x)$. In fact, the 
equations (\ref{3e}), (\ref{3ap}) and (\ref{3h}) are closely related 
to the recursion relations for spherical Bessel functions 
\cite{Abramowitz}. 

The solutions (\ref{3ak}) fulfill \cite{Gradstein}
\beq\label{3q}
(\varphi_\nu, \varphi_\mu)_{L_2} 
= {2 \e^{- \imu \pi (\nu^* - \mu) / 2} \over \nu^* + \mu + 1}
{\sin \pi (\nu^* - \mu) \over \pi (\nu^* - \mu)} 
\eeq
for $\re (\nu^* + \mu) > - 1$. Here the value of $\varphi_\nu(-x)$, 
$x > 0$, has been chosen above the cut, i.~e. $- x = \e^{\imu \pi} x$. 
This result has quite interesting consequences: First of all, for 
integer $n \geq 0$ only $\varphi_n$, but not $\varphi_{- n - 1}$ is 
$L_2$ normalizable. Furthermore, the functions
\beq
\hat{\varphi}_n(x) 
= \sqrt{n + 1 / 2} \, \varphi_n(x) 
= \sqrt{2 n + 1 \over \pi} j_n(x) \,,\quad n \geq 0 \,,
\eeq
are orthonormal in $L_2$:
\beq
(\hat{\varphi}_n, \hat{\varphi}_m)_{L_2} = \delta_{n m} \quad,
\eeq
in other words, the linear transformation $T: \hat{\mcal{H}} \to L_2$
defined by $T \hat{\psi}_n = \hat{\varphi}_n$ is an isometry. But it 
is not a unitary transformation, because the $\hat{\varphi}_n$ do 
{\em not\/} span the whole space $L_2$. This can be seen most easily 
by considering
\beq
D(x, y) = \sum_{n = 0}^\infty \hat{\varphi}_n(x) \hat{\varphi}_n^*(y)
= \sum_{n = 0}^\infty {2 n + 1 \over \pi} j_n(x) j_n(y) \quad.
\eeq
Completeness of the $\hat{\varphi}_n$ would require 
$D(x, y) = \delta(x - y)$, but instead one obtains \cite{Abramowitz}
\beq
D(x, x) = {1 \over \pi} \quad.
\eeq
Even worse, the formally $L_2$ Hermitian Hamiltonian (\ref{3p}) is not
even Hermitian: It follows from (\ref{3q}) that
\beq
\|\varphi_\nu\|_{L_2}^2
= {\e^{- \pi \im \nu} \over \re \nu + 1 / 2} 
{\sinh 2 \pi \im \nu \over 2 \pi \im \nu} \quad \mbox{for } 
\re \nu > - {1 \over 2} \quad,
\eeq
thus $H_{L_2}$ has eigenfunctions with complex eigenvalues 
$E_\nu =(\nu + 1 / 2)^2 / 2 \in {\Bbb C}$ in $L_2$, so that
$(\varphi_\nu, H_{L_2} \varphi_\nu)_{L_2} 
= E_\nu \|\varphi_\nu\|_{L_2}^2 
\neq E_\nu^* \|\varphi_\nu\|_{L_2}^2
= (H_{L_2} \varphi_\nu, \varphi_\nu)_{L_2}$.

Our initial aim was to apply the uniqueness theorem to the canonical
formulation (\ref{3am}). But we have not checked whether the canonical
operators $\hat{x}^{\mrm{c}}$ and $\hat{p}^{\mrm{c}}$ are self-adjoint
(as required by the uniqueness theorem). Although the Hermiticity and
dense domain of definition of these operators are a necessary
condition for this \cite{Prugovecki}, it is in general not an easy
task to check the self-adjointness. In our case, a simple argument can
be given that shows that no unitary transformation
$T: \hat{\mcal{H}} \to L_2$ that maps $\hat{x}^{\mrm{c}}$ and
$\hat{p}^{\mrm{c}}$ to the canonical position and momentum operators 
of the space $L_2$,
$x^{\mrm{c}}_{L_2} = T \hat{x}^{\mrm{c}} T^{- 1}$,
$p^{\mrm{c}}_{L_2} = T \hat{p}^{\mrm{c}} T^{- 1}$, exists: The 
operator $\hat{p}^{\mrm{c}}$ is bounded, from (\ref{3h}) one finds
\beq
\|\hat{p}^{\mrm{c}} \psi\|_{\hat{\mcal{H}}} \leq \sqrt{3 \over 5} \,
\|\psi\|_{\hat{\mcal{H}}} \quad \mbox{for all } 
\psi \in \hat{\mcal{H}} \quad.
\eeq
Under a unitary transformation $T$ the boundedness properties do not
change, hence the bounded operator $\hat{p}^{\mrm{c}}$ cannot be
mapped to the unbounded operator $p^{\mrm{c}}_{L_2}$.

This argument can even be generalized: The operator 
$\hat{p}^{\mrm{c}}$ in (\ref{3i}) is not the only momentum operator 
that can be assigned to $\hat{x}^{\mrm{c}}$, because if
$f(\hat{x}^{\mrm{c}})$ is a Hermitian function of $\hat{x}^{\mrm{c}}$,
then
\beq\label{3a}
\pi^{\mrm{c}} = \hat{p}^{\mrm{c}} + f(\hat{x}^{\mrm{c}})
\eeq
is also a Hermitian operator with
$[\hat{x}^{\mrm{c}}, \pi^{\mrm{c}}] = \imu$. Now suppose that this set
$(\hat{x}^{\mrm{c}}, \pi^{\mrm{c}})$ can be mapped to the canonical
position and momentum operators of $L_2$ by a unitary transformation
$T$. Then for $\varphi \in L_2$, \eq{3a} gives
\begin{eqnarray}
(\hat{p}^{\mrm{c}}_{L_2} \varphi)(x)
& = & (T \hat{p}^{\mrm{c}} T^{- 1} \varphi)(x)
= - \imu {\md \varphi \over \md x} - f(x) \varphi(x) \\
& = & - \imu \e^{\imu F(x)} {\md \over \md x} 
\left( \e^{- \imu F(x)} \varphi(x) \right) \\
& = & (U p^{\mrm{c}}_{L_2} U^{- 1} \varphi)(x) \quad,
\end{eqnarray}
where $\md F / \md x = f(x)$, and where the unitary operator $U: L_2
\to L_2$ is the gauge transformation
$U = \exp \imu  F(x^{\mrm{c}}_{L_2})$. Thus the canonical $L_2$ 
momentum would be given by
\beq
p^{\mrm{c}}_{L_2} = U^{- 1} \hat{p}^{\mrm{c}}_{L_2} U \quad.
\eeq
Here all operators on the right-hand side are bounded, in 
contradistinction to the fact that $p^{\mrm{c}}_{L_2}$ is not bounded. 
This shows that if $\hat{p}^{\mrm{c}}$ is bounded, then also all 
operators $\pi^{\mrm{c}}$ of the form (\ref{3a}) cannot be unitarily 
equivalent to the canonical $L_2$ momentum.

Physically this means that one cannot interpret $\hat{x}^{\mrm{c}}$
and $\hat{p}^{\mrm{c}}$ or $\pi^{\mrm{c}}$ as position and momentum 
operators, resp. Thus our conclusion is that the physical 
interpretation of the model of Cannata et al.\ is still unclear. Of 
course, it may be possible that other, self-adjoint canonical 
operators can be found. As we have already pointed out, such a 
formulation is, in our opinion, necessary for the very definition of 
the model.

\subsection{The model of Hatano and Nelson}
\label{s33}

In the context of studies of delocalization phenomena, the model of 
Hatano and Nelson \cite{Hatano...} has attracted a lot of interest 
recently \cite{Nelson,Feinberg}. It is defined in one dimension by the 
non-Hermitian Hamiltonian
\beq\label{3t}
H(g) = {(p + \imu g)^2 \over 2 m} + V(x) \quad,
\eeq
where $g$ is a real parameter connected to an externally applied
magnetic field, and $V(x)$ is a random potential. It has been 
demonstrated numerically that at a certain critical value 
$g = g_{\mrm{c}}$ a localized wave function turns into a delocalized 
one, and it has been suggested that this behaviour signals the 
occurrence of a delocalization phase transition. 

The numerical demonstration of a delocalization transition at the
critical value $g_{\mrm{c}}$ is based on the use of a modified scalar
product in the space of quantum-mechanical states \cite{Hatano...}. 
There has been some controversy about this point \cite{Silvestrov}, 
but at present a consensus seems to have been reached that the 
delocalization transition is only visible if the density distribution 
of particles is calculated according to a scalar product based on the 
product of left- and right-eigenfunctions of the Hamiltonian 
(\ref{3t}) \cite{Hatano}.

We want to point out here that a physical interpretation that is based
on this modified scalar product is only consistent if the underlying
Hilbert space $\mcal{H}$ is chosen accordingly. In fact, as we will 
show, this Hilbert space is just of the kind considered in 
Section~\ref{s2}, i.~e.\ below $g_{\mrm{c}}$ the Hamiltonian 
(\ref{3t}) is Hermitian in $\mcal{H}$. But it turns out that under 
reasonable conditions this Hilbert space explicitly depends on the 
parameter $g$, so that the $g$ dependence of the Hamiltonian 
(\ref{3t}) is not the {\em complete\/} dependence of the model of
Hatano and Nelson on $g$. 

It is the aim of this Section to investigate this dependence. This is 
important for the phenomenological implications of the model of Hatano 
and Nelson, because $g$ is considered to be an external parameter that 
can be varied. 

We want to emphasize that we are dealing here only with the 
{\em quantum-mechanical\/} model of \cite{Hatano...}, not with its 
applications to statistical-mechanical problems like vortex depinning 
in type-II superconductors or population biology \cite{Nelson}.

Consider a set of eigenfunctions $\psi_n(x; g)$ of (\ref{3t}),
\beq\label{3u}
H(g) \psi_n(x; g) = E_n(g) \psi_n(x; g) \quad,
\eeq
determined as solutions of the corresponding differential equation and 
being elements of a Hilbert space $\mcal{H}_0$ (e.~g.\ the space
$L_2(- \infty, \infty)$ or the space of periodic functions). In 
\cite{Hatano...}, examples are given in which the eigenvalues $E_n(g)$ 
are real for $g$ being smaller than some critical value $g_{\mrm{c}}$, 
for $g > g_{\mrm{c}}$ they become complex.

Let us first recall the construction of the modified scalar product as
introduced by Hatano and Nelson. We refer here to the detailed
description given in \cite{Hatano}, according to which the $\psi_n$ 
can be chosen such that
\beq\label{3v}
\int \md x \, \psi_n(x; - g)^* \psi_m(x; g) 
= (\psi_n(- g), \psi_m(g))_{\mcal{H}_0}
\eeq
is always well-defined and given by
\beq\label{3w}
(\psi_n(- g), \psi_m(g))_{\mcal{H}_0} = \delta_{n m} \quad.
\eeq
The probabilistic interpretation of the model is based on this
bilinear form; the integrand in (\ref{3v}) for $n = m$,
$\varrho(x) = \psi_n(x; - g)^* \psi_n(x; g)
\equiv \psi_n^{\mrm{L}}(x; g) \psi_n^{\mrm{R}}(x; g)$ (in the notation 
of \cite{Hatano}, the superscripts L and R denoting left- and 
right-eigenfunctions) is interpreted as the particle density in the 
bulk of the sample. 

It is important to realize that (\ref{3w}) rests upon some quite 
non-trivial assumptions. First of all, as the $\mcal{H}_0$ adjoint of 
$H(g)$ is given by $H^\dagger(g) = H(- g)$, \eq{3w} can only hold 
provided the spectra of $H(g)$ and $H(- g)$ can be ordered in such a 
way that \cite{Hatano}
\beq\label{3x}
E_n(- g)^* = E_n(g) \quad.
\eeq
By taking the complex conjugate of the differential equation 
corresponding to the Schr\"odinger equation (\ref{3u}), we see that if 
$H(g) \psi_n(x; g) = E_n(g) \psi_n(x; g)$, then
\beq
H(g) \psi_n^*(x; g) = E_n^*(g) \psi_n^*(x; g) \quad,
\eeq
i.~e.\ the eigenvalues come in complex conjugate pairs. Hence 
(\ref{3x}) asserts that the spectra of $H(g)$ and $H(- g)$ coincide.

In order to construct a new Hilbert space that is based on a scalar
product consistent with (\ref{3w}), we again start with the 
infinite-dimensional vector space 
$\mcal{V}(g) = \vspan \{\psi_n(g), n = 0, 1, \ldots\}$ of the finite 
superpositions of the eigenfunctions. Next, we define a linear 
operator $M_g: \mcal{V}(g) \to \mcal{V}(- g)$ which acts as
\beq\label{3y}
M_g \psi_n(g) = \psi_n(- g) 
\eeq
and use it to define a scalar product in $\mcal{V}(g)$ through
\beq
(\psi, \psi')_{\mcal{V}(g)} \equiv (M_g \psi, \psi')_{\mcal{H}_0} 
\quad.
\eeq
Proceeding as in Section~\ref{s2}, we can complete this space with
respect to the norm 
$\|\psi\|_{\mcal{V}(g)} = \sqrt{(\psi, \psi)_{\mcal{V}(g)}}$ to obtain 
the separable Hilbert space $\mcal{H}$. Then, by (\ref{3w}), the 
eigenfunctions $\psi_n(g)$ form an orthonormal basis in $\mcal{H}$. 

There may also be cases in which one is considering an enumerable set
of solutions $\psi_n(x; g)$ of (\ref{3u}) that are not elements of a
space $\mcal{H}_0$ (e.~g.\ because they are not normalizable in
$\mcal{H}_0$). One can then still construct $\mcal{H}$ in complete
analogy to Section~\ref{s2} by considering this set of solutions 
$\psi_n(x; g)$, defining $\mcal{V}(g)$ as above, and defining the 
scalar product in $\mcal{V}(g)$ so as to fulfill
\beq
(\psi_n(g), \psi_m(g))_\mcal{H} = \delta_{n m} \quad.
\eeq
This definition is always possible; it does not rely on additional 
assumptions like (\ref{3x}). Condition (\ref{3x}) only has to do with 
the explicit form of the metric operator $M_g$ relating the scalar 
products in $\mcal{H}$ and $\mcal{H}_0$.

We can now formulate the model of Hatano and Nelson entirely in the
space $\mcal{H}$, and since the spectrum is real for 
$g < g_{\mrm{c}}$, we find as in (\ref{2o}) 
\beq\label{3z}
H^\ddagger(g) = H(g) \quad \mbox{ for } g < g_{\mrm{c}} \quad.
\eeq

We emphasize that it is essential for the following discussion that 
the metric operator $M_g$ in (\ref{3y}) depends on $g$, because then 
it may happen that the space $\mcal{H}$ will be different for 
different values of $g$.

For matrix elements of an operator $A_\mcal{H}$ and vectors $\psi$, 
$\psi'$ that are contained both in $\mcal{H}$ and $\mcal{H}_0$, and 
for which both the Hermitian adjoint $A_\mcal{H}^\dagger$ in 
$\mcal{H}_0$ and $A_\mcal{H}^\ddagger$ in $\mcal{H}$ are defined, one 
finds
\begin{eqnarray}
\lefteqn{(A_\mcal{H}^\ddagger \psi, \psi')_\mcal{H}
= (\psi, A_\mcal{H} \psi')_\mcal{H}
= (M_g \psi, A_\mcal{H} \psi')_{\mcal{H}_0}} \nonumber \\
& = & (A_\mcal{H}^\dagger M_g \psi, \psi')_{\mcal{H}_0}
= (M_g M_{- g} A_\mcal{H}^\dagger M_g \psi, \psi')_{\mcal{H}_0} 
\nonumber \\
& = & (M_{- g} A_\mcal{H}^\dagger M_g \psi, \psi')_\mcal{H} 
\end{eqnarray}
since $M_g M_{- g} = 1$, so that
\beq
A_\mcal{H}^\ddagger = M_{- g} A_\mcal{H}^\dagger M_g \quad.
\eeq

Although $\mcal{H}$ is well-defined by the construction outlined 
above, we do not know much about it as the eigenfunctions 
$\psi_n(x; g)$ are not explicitly known. (To our knowledge, no 
non-trivial, exactly solvable model of the form (\ref{3t}) that 
exhibits a delocalization transition exists.) One therefore has to 
rely on some assumptions, and we will assume here that the operator 
$x$ in (\ref{3t}) is $\mcal{H}$ Hermitian,
\beq
x^\ddagger = x \quad.
\eeq
This assumption is quite natural, because according to 
\cite{Hatano...} the fact that a delocalization transition (leading 
from a real spectrum to a complex one) occurs, does not depend on the 
detailed form of the potential $V(x)$. Since below $g_{\mrm{c}}$ the 
Hamiltonian $H(g)$ is Hermitian (cf.\ (\ref{3z})), and since this has 
to be true for a large class of potentials, the operator $x$ should be 
Hermitian. Note that here, contrary to the treatment of the examples 
in Sections~\ref{s31} and \ref{s32}, we treat $x$ as a physical 
observable. 

Consider the case $g < g_{\mrm{c}}$: We then have from the Hermiticity 
of $H(g)$ and $x$
\beq
(p^\ddagger - \imu g)^2 = (p + \imu g)^2 \quad.
\eeq
This equation has two solutions:
\begin{eqnarray}
p^\ddagger & = & - p \quad \mbox{and} \label{3ab} \\
p^\ddagger & = & p + 2 \imu g \quad. \label{3ac}
\end{eqnarray}

The first solution (\ref{3ab}) would be compatible with an $M_g$ that 
is {\em independent\/} of $g$, since $M_{- g} p M_g = - p$ in this 
case. The second solution (\ref{3ac}),
\beq\label{3ai}
M_{- g} p M_g = p + 2 \imu g \quad,
\eeq
requires a $g$ dependent $M_g$. But the first solution is physically
unacceptable. The reason is that (as $H(g)$ is Hermitian) we can 
assume that the time evolution is governed by the Schr\"odinger
equation, so that the expectation value 
$\langle x \rangle(t) = (\psi(t), x \psi(t))_\mcal{H}$ evolves 
according to
\beq\label{3aj}
{\md \over \md t} \langle x \rangle(t) 
= \imu \langle [H(g), x] \rangle(t)
= \langle v \rangle(t)
\eeq
with $v = (p + \imu g) / m$. Hence $p + \imu g \equiv p^{\mrm{kin}}$ 
is the kinetic momentum. (This observable has also been considered in 
\cite{Hatano...}.) The first solution (\ref{3ab}) then corresponds to 
an anti-Hermitian velocity operator, $v^\ddagger = - v$, giving a 
purely imaginary expectation value for the velocity. Only the second 
solution (\ref{3ac}) leads to a real expectation value for the 
velocity. Therefore, we only consider (\ref{3ac}) to be physically 
meaningful, and consequently $M_g$ has to depend on $g$.

We now want to discuss how the quantum-mechanical predictions of the 
model vary with $g$. The situation in quantum mechanics is similar to 
the situation in classical Hamiltonian mechanics: There one has two 
canonically conjugate variables $x^{\mrm{c}}$ and $p^{\mrm{c}}$ which 
are the independent variables, and the Hamiltonian has to be expressed 
in terms of these two variables:
\beq\label{3ad}
H = H(g; x^{\mrm{c}}, p^{\mrm{c}}) \quad.
\eeq
Varying any parameter $g$ appearing in the Hamiltonian, the system
reacts in precisely the way given by the $g$ dependence of the
formulation (\ref{3ad}). In the quantum-mechanical case, these two 
canonically conjugate variables $x^{\mrm{c}}$ and $p^{\mrm{c}}$ have 
to be two operators fulfilling canonical commutation relations
$[x^{\mrm{c}}, p^{\mrm{c}}] = \imu$ and have to be self-adjoint. By 
this last requirement, the structure of the underlying Hilbert space 
comes into play. Having such a formulation, we can use the uniqueness 
theorem to map $H(g; x^{\mrm{c}}, p^{\mrm{c}})$ uniquely (modulo 
unitary equivalence) to the $L_2$ formulation 
$H_{L_2} = H(g; x^{\mrm{c}}_{L_2}, p^{\mrm{c}}_{L_2})$. Thus, the $g$ 
dependence of $H(g; x^{\mrm{c}}_{L_2}, p^{\mrm{c}}_{L_2})$ describes 
the complete $g$ dependence of the physical system.

Applying this reasoning to the model of Hatano and Nelson, one has to 
find the two canonically conjugate operators $x^{\mrm{c}}$ and 
$p^{\mrm{c}}$. Again, we assume that $x$ is Hermitian in $\mcal{H}$,
whereas $p^{\mrm{c}}$ is non-Hermitian, see (\ref{3ac}). But as 
$[x, p] = \imu$, one finds that one may choose
\beq\label{3ae}
x^{\mrm{c}} = x \quad \mbox{and} \quad
p^{\mrm{c}} = {1 \over 2} (p + p^\ddagger) 
\eeq
as the set of Hermitian operators with canonical commutation 
relations. (For $g < g_{\mrm{c}}$, \eq{3ae} and (\ref{3ac}) just give 
$p^{\mrm{c}} = p + \imu g = p^{\mrm{kin}}$.) 

The next task is then to 
express the Hamiltonian (\ref{3t}) through $x^{\mrm{c}}$ and 
$p^{\mrm{c}}$. In order to do this, one has to know how $p^\ddagger$ 
acts in $\mcal{H}$ for arbitrary $g$. Again, this would be clear once 
we knew the states $\psi_n$ explicitly. But since we don't know them, 
we have to rely on some assumptions, and here we will restrict our 
discussion to a physically rather uninteresting case, namely the case 
of wave functions that are defined with infinite, non-periodic 
boundary conditions. It is well-known \cite{Hatano...} that under 
these conditions there is no $g$ dependence at all. The numerical 
demonstrations of delocalization transitions have always been carried 
out for wave functions with periodic boundary conditions. Still, the 
example of infinite, non-periodic boundary conditions serves 
illustrative purposes. At the end, we will comment on the case of 
periodic boundary conditions.

The point is that for non-periodic boundary conditions the unitary
transformation $T_g: \mcal{H} \to L_2$ is known to be only a function 
of $x = x^{\mrm{c}}$ \cite{Hatano...}:
\beq
T_g = T_g(x^{\mrm{c}}) \quad.
\eeq
Such a $T_g$ acts on states $\psi \in \mcal{H}$ simply as
$(T_g(x^{\mrm{c}}) \psi)(x) = T_g(x) \psi(x)$, so that
\beq
(\psi, \psi')_\mcal{H} = (T_g \psi, T_g \psi')_{L_2} 
= \int \md x \, |T_g(x)|^2 \psi^*(x) \psi'(x) \quad.
\eeq
From this equation one can deduce how $p^\ddagger$ acts on states of 
$\mcal{H}$: A simple calculation based on
$(p^\ddagger \psi, \psi')_\mcal{H} = (\psi, p \psi')_\mcal{H}$ and 
partial integration shows that
\beq
p^\ddagger = p - \imu {\md \over \md x^{\mrm{c}}} 
\ln (T_g T_g^\ddagger) \quad.
\eeq
Thus, from (\ref{3ae}) we obtain
\beq
p = p^{\mrm{c}} + {\imu \over 2} {\md \over \md x^{\mrm{c}}} 
\ln (T_g T_g^\ddagger) \quad.
\eeq
Substituting this into (\ref{3t}) leads to the expression
\begin{eqnarray}
\lefteqn{H(g) 
= H(g; x^{\mrm{c}}, p^{\mrm{c}})} \nonumber \\
& = & {1 \over 2 m} \left( p^{\mrm{c}} + {\imu \over 2}
{\md \over \md x^{\mrm{c}}} \ln (T_g T_g^\ddagger)
+ \imu g \right)^2 + V(x^{\mrm{c}}) \quad. \label{3af} 
\end{eqnarray}
Equation (\ref{3af}) represents the canonical form of the Hamiltonian 
of the model of Hatano and Nelson for infinite, non-periodic boundary
conditions. As before, in this form the model can immediately be 
formulated in the space $L_2$ by just replacing $x^{\mrm{c}}$, 
$p^{\mrm{c}}$ by $x_{L_2}^{\mrm{c}}$, $p_{L_2}^{\mrm{c}}$, resp. As 
one can see, we have canonically coupled a nontrivial, purely 
imaginary field
\beq
A(x; g) = {\imu \over 2} {\md \over \md x} \ln |T_g(x)|^2 + \imu g
\eeq
to the system with $g = 0$.

The spectrum of (\ref{3af}) is real for $g < g_{\mrm{c}}$ and complex 
for $g > g_{\mrm{c}}$. This means that 
$H(g; x^{\mrm{c}}, p^{\mrm{c}})$ is Hermitian below $g_{\mrm{c}}$ and 
non-Hermitian above $g_{\mrm{c}}$ (see (\ref{3z})). The Hermiticity 
translates into a vanishing anti-commutator
$\{p^{\mrm{c}}, A(x^{\mrm{c}}; g)\} = 0$ for $g < g_{\mrm{c}}$. 
Formulating this in $L_2$,
\beq
\{- \imu \md / \md x, A(x; g)\} \varphi(x) = 0 \quad \mbox{for all }
\varphi \in L_2 \quad,
\eeq
leads to the equation
$\varphi (\md A / \md x) = - 2 A (\md \varphi / \md x)$ which---as it 
has to hold for all $\varphi \in L_2$---requires
\beq
A(x; g) = 0 \quad (\mbox{for } g < g_{\mrm{c}}) \quad.
\eeq
For $g > g_{\mrm{c}}$ the spectrum becomes complex, and 
$A(x; g) \neq 0$ is possible. Hence one can write explicitly
\beq\label{3ag}
H(g; x^{\mrm{c}}, p^{\mrm{c}}) 
= {1 \over 2m} [p^{\mrm{c}} + \Theta(g - g_{\mrm{c}}) 
A(x^{\mrm{c}}; g)]^2 + V(x^{\mrm{c}}) \quad. 
\eeq

In fact, for infinite, non-periodic boundary conditions it is 
well-known \cite{Hatano...} that $T_g$ is given by
\beq\label{3ah}
T_g(x^{\mrm{c}}) = \exp (- g x^{\mrm{c}}) \quad,
\eeq
leading to
\beq
A(x; g) = 0 \quad \mbox{for all } g \quad,
\eeq
i.~e.\ $g_{\mrm{c}} = \infty$ in (\ref{3ag}), and the spectrum is real 
for all $g$. 

The transformation $T_g$ in (\ref{3ah}) is often refered to as an 
``imaginary gauge transformation''. Our analysis shows that this 
notion is quite misleading: Nothing has been {\em gauged away\/} by 
$T_g$; we just have two formulations of the same physics, realized in 
the two different, though unitarily equivalent Hilbert spaces 
$\mcal{H}$ and $L_2$.

The problem in the case of Hilbert spaces $\mcal{H}_0$ and $\mcal{H}$
that contain wave functions with periodic boundary conditions is that 
in these spaces only the differentiation operator $p$ is defined; the
operator $x$ has a vanishing domain of definition. This makes it 
impossible to construct a canonical formulation for which the 
uniqueness theorem may be used. Still, if one chooses a periodic
potential in (\ref{3t}), our reasoning applies up to \eq{3ai}; only
the argument given in (\ref{3aj}) can---strictly speaking---no longer 
be used. This means that also in the case of periodic boundary 
conditions it may happen that the space $\mcal{H}$ is an unknown 
function of $g$, making it a priori impossible to compare the 
predictions of the model for different values of $g$. 

To our knowledge, this aspect has not been discussed in the literature 
up to now, but it is in our opinion of fundamental importance for the 
phenomenological interpretation of the model of Hatano and Nelson. We 
see, for example, that the canonical formulation (\ref{3ag}) describes 
a {\em non-continuously differentiable change of the Hamiltonian\/} at 
the transition point $g_{\mrm{c}}$. Whether such a behaviour should be 
called a phase transition is not clear to us; it seems to be more 
similar to the sudden switching-on of a magnetic field as the 
parameter $g$ passes $g_{\mrm{c}}$. It is clear that the behaviour of 
the system will then change, localized states may become delocalized, 
but one would hardly call this a phase transition. 

Let us mention another aspect of the model of Hatano and Nelson. We
have seen in (\ref{3z}) that as long as the spectrum of $H(g)$ is 
real, the non-Hermiticity of $H(g)$ with respect to the space 
$\mcal{H}_0$ is a rather superficial one that can be avoided by
considering the space $\mcal{H}$. Thus the time evolution can 
consistently be described by the Schr\"odinger equation, leading to a 
unitary time-evolution operator. The situation changes when the 
spectrum becomes complex. It is then no longer possible to construct a 
Hilbert space in which the Hamiltonian is Hermitian and which at the 
same time contains eigenstates with complex eigenvalues. Although a 
completely consistent probabilistic interpretation is still possible, 
the interpretation of the Hamiltonian as the generator of the time 
evolution would then lead to a non-unitary time-evolution operator 
with all the problems mentioned in Section~\ref{s21}. 

\section{Conclusions}
\label{s4}

In this work we have studied the quantum-mechanical interpretation of
models with non-Hermitian Hamiltonians (in the usual sense of the 
space $L_2$) and real spectra. 

Assuming that the systems under consideration are closed, and that 
the Hamiltonians are the generators of the time evolution for these 
systems, we construct separable Hilbert spaces in which the 
Hamiltonians are Hermitian. 

Within this construction, we set up a canonical formulation in which 
the Hamiltonian is only a function of two Hermitian operators 
$x^{\mrm{c}}$ and $p^{\mrm{c}}$ that fulfill canonical commutation 
relations. Then a unique physical interpretation is obtained if these
operators are self-adjoint, because in this case the model can be 
formulated as a Hermitian problem in the space $L_2$. If, on the other 
hand, this equivalent $L_2$ description cannot be achieved, we 
consider the model to be quantum-mechanically inconsistent. 

We apply the above construction to a number of models recently
discussed in the literature.

As a first example, we have analysed a simple non-Hermitian model that 
is based on the Hermitian one-dimensional Coulomb problem on the real 
half axis. Motivated by the approach of Bender et~al.\ 
\cite{Bender...}, we extend the eigenfunctions to complex values of 
the coordinate. We then show that the canonical formulation of this
example does not contain any new information compared to the original, 
Hermitian version of the model. 

The next example is the model introduced by Cannata et~al.\ 
\cite{Cannata}. Although here we also find a canonical formulation, it 
turns out that no equivalent $L_2$ formulation exists. In our opinion 
this may indicate that the model is intrinsically inconsistent.

Our last example is the phenomenologically important model of Hatano 
and Nelson \cite{Hatano...}. Here the appropriate Hilbert space is 
already determined by the introduction of the modified scalar product 
in \cite{Hatano...}. Under these conditions, the dependence of the 
Hamiltonian on the external parameter $g$ that is responsible for the 
non-Hermiticity of the Hamiltonian does not fully describe how the 
system varies with $g$. For the special case of wave functions with 
infinite, non-periodic boundary conditions, we give the canonical 
$L_2$ formulation which clearly exhibits this complete $g$ dependence. 
It turns out that at the critical value $g = g_{\mrm{c}}$, where the 
states of the system undergo a delocalization transition, also the 
$L_2$ version of the Hamiltonian changes in a non-continuously 
differentiable way from Hermitian to non-Hermitian. In our opinion, 
this makes the usual interpretation of this transition as a phase 
transition questionable. We argue that in the phenomenologically 
relevant case with periodic boundary conditions similar effects may 
also occur.

\section*{Acknowledgment}
L. S. acknowledges support of Deutsche Forschungsgemeinschaft and the
warm hospitality at Regensburg University.

\appendix

\section{Properties of the canonical variables for the model of 
Cannata et al.}
\label{sa}

In (\ref{3o}) the operator $\hat{x}^{\mrm{c}}$ is defined on the 
subset $\{\xi_1, \xi_2, \ldots\}$ which is dense in $\hat{\mcal{H}}$.
\eq{3an} shows that $\hat{x}^{\mrm{c}}$ is Hermitian. We can therefore
use the fact that every densely defined, Hermitian operator has a
closed extension \cite{Prugovecki}, to extend the domain of definition
of $\hat{x}^{\mrm{c}}$ to some larger space than $\hat{\mcal{V}}_\xi$.
The problem is that $\hat{x}^{\mrm{c}} \hat{\psi}_n$ is not defined by
this procedure (cf.\ (\ref{3ao})). We will therefore in the following
define every operator on the space $\hat{\mcal{V}}_\xi$, check its 
algebraic properties there, and will then implicitly assume that we
consider closed extensions where this is possible.

To start, we rewrite (\ref{3e}) and (\ref{3ap}) with the help of
(\ref{3aq}) and (\ref{3r}) in the form
\beq
\e^{- \imu x} \xi_n 
= {\xi_{n - 1} \over 2 (n - 1 + 1 / 2)} 
+ {\xi_{n + 1} \over 2 (n + 1 + 1 / 2)}
\eeq
for $n \geq 2$, and 
$\e^{- \imu x} \xi_1 
= \xi_2 / 5 - \sum_{k = 1}^\infty (- \imu)^k \xi_k$. It follows that 
although
\beq\label{aa}
\e^{- \imu x} \hat{x}^{\mrm{c}} \xi_n = \xi_n \quad \mbox{for all } 
n \geq 1 \quad,
\eeq
one obtains
\beq\label{ae}
\hat{x}^{\mrm{c}} \e^{- \imu x} \xi_n = \xi_n \quad \mbox{ only for }
n \geq 2 \quad,
\eeq
whereas $\hat{x}^{\mrm{c}} \e^{- \imu x} \xi_1$ is divergent.

In order to construct the canonically conjugate momentum operator
$\hat{p}^{\mrm{c}}$, we first note that (\ref{3d}) leads to
\beq\label{ab}
\e^{- \imu x} p \hat{\psi}_n 
= {\sqrt{n + 1 / 2} \over 2} \left( \!
{\hat{\psi}_{n - 1} \over \sqrt{n - 1+ 1 / 2}} 
- {\hat{\psi}_{n + 1} \over \sqrt{n + 1+ 1 / 2}} \! \right)
\eeq
for $n \geq 1$ and 
$\e^{- \imu x} p \hat{\psi}_0 
= \imu \hat{\psi}_0 / 2 - \hat{\psi}_1 / (2 \sqrt{3})$. Then we use 
$p \, \e^{- \imu x} - \e^{- \imu x} p = - \e^{- \imu x}$ together with 
(\ref{aa}) to get $p = \e^{- \imu x} p \hat{x}^{\mrm{c}} - 1$. 
Combining this with (\ref{ab}) gives for all $n \geq 1$
\beq
p \xi_n 
= (n - 1 + 1 / 2) \xi_n 
- (2 n + 1) {\hat{\psi}_{n + 1} \over \sqrt{n + 1 + 1 / 2}} \quad.
\eeq
This equation shows that $p$ is anti-Hermitian:
\beq
(\xi_n, p \xi_m)_{\hat{\mcal{H}}} 
= \delta_{n + 2, m} - \delta_{n, m + 2} 
= - (p \xi_n, \xi_m)_{\hat{\mcal{H}}}
\eeq
for all $n, m \geq 1$. Now define
\beq
\hat{p}^{\mrm{c}} 
= - {\imu \over 2} (\e^{- \imu x} p + p \, \e^{- \imu x})
= - \imu \e^{- \imu x} (p - 1 / 2) \quad.
\eeq
This operator fulfills
\beq
\hat{p}^{\mrm{c}} \xi_n 
= {\imu \over 2} \left( - {n - 1 \over n - 1 + 1 / 2} \xi_{n - 1} 
+ {n + 2 \over n + 1 + 1 / 2} \xi_{n + 1} \right)
\eeq
for all $n \geq 1$. In the matrix element 
$(\xi_n, \hat{p}^{\mrm{c}} \xi_m)_{\hat{\mcal{H}}} 
= - \imu (\xi_n, \e^{- \imu x} (p - 1 / 2) \xi_m)_{\hat{\mcal{H}}}$ 
(with $n, m \geq 1$) the vector
\beq
(p - 1 / 2) \xi_m 
= (m - 1) \xi_m - (2 m + 1) {\hat{\psi}_{m + 1} 
\over \sqrt{m + 1 + 1 / 2}}
\eeq
never has a component proportional to $\hat{\psi}_0$ (see 
(\ref{3aq})), because for $m = 1$ the first term vanishes. Then, as a 
consequence of (\ref{3ar}), the operator $\e^{- \imu x}$ in 
$(\xi_n, \hat{p}^{\mrm{c}} \xi_m)_{\hat{\mcal{H}}}$ can be shifted to 
the left side for all values $n, m \geq 1$. This, together with the 
anti-Hermiticity of $p$ shows that $\hat{p}^{\mrm{c}}$ is Hermitian in 
the entire space $\hat{\mcal{V}}_\xi$.

The commutator $[\hat{x}^{\mrm{c}}, \hat{p}^{\mrm{c}}]$ can be
calculated in a similar way: One has 
$\hat{x}^{\mrm{c}} \hat{p}^{\mrm{c}} \xi_n 
= - \imu \hat{x}^{\mrm{c}} \e^{- \imu x} (p - 1 / 2) \xi_n$. Since 
$(p - 1 / 2) \xi_n$ never has a $\xi_1$ component, one can apply 
(\ref{ae}), so that
\beq\label{af}
\hat{x}^{\mrm{c}} \hat{p}^{\mrm{c}} \xi_n
= - \imu (p - 1 / 2) \xi_n \quad \mbox{for all } n \geq 1 \quad.
\eeq
Writing $\hat{p}^{\mrm{c}} = - \imu (p + 1 / 2) \e^{- \imu x}$, one
can use (\ref{aa}) to derive 
$\hat{p}^{\mrm{c}} \hat{x}^{\mrm{c}} \xi_n 
= - \imu (p + 1 / 2) \xi_n$ for all $n \geq 1$. Combining these 
relations leads to the desired result:
\beq
[\hat{x}^{\mrm{c}}, \hat{p}^{\mrm{c}}] \xi_n = \imu \xi_n \quad
\mbox{for all } n \geq 1 \quad.
\eeq
In summary, the operators $\hat{x}^{\mrm{c}}$ and $\hat{p}^{\mrm{c}}$
are Hermitian, densely defined operators in $\hat{\mcal{H}}$ that 
fulfill canonical commutation relations.

One can now express the Hamiltonian (\ref{3f}) as a function of
$\hat{x}^{\mrm{c}}$ and $\hat{p}^{\mrm{c}}$: If we restrict the
discussion for the moment to the subspace 
$\vspan \{\xi_2, \xi_3, \ldots\}$, we can replace $\e^{2 \imu x}$ in 
(\ref{3f}) by $(\hat{x}^{\mrm{c}})^2$. Then (\ref{af}) gives
\beq\label{ac}
H = {1 \over 2} \left[ \left( \imu \hat{x}^{\mrm{c}} \hat{p}^{\mrm{c}}
+ {1 \over 2} \right)^2 + (\hat{x}^{\mrm{c}})^2 \right] \quad.
\eeq
Applying $\xi_n$ to it, one should note that, for example,
$(\hat{x}^{\mrm{c}})^2 \xi_n 
= 2 \sqrt{n + 1 / 2} \, \hat{x}^{\mrm{c}} \hat{\psi}_n$ is not 
defined. But the terms on the right-hand side of (\ref{ac}) can be 
arranged in the following way:
\beq\label{ad}
H 
= {1 \over 2} \left[ \hat{x}^{\mrm{c}} \left[ \hat{x}^{\mrm{c}} 
\left( 1 - (\hat{p}^{\mrm{c}})^2 \right) + 2 \imu \hat{p}^{\mrm{c}} 
\right] + {1 \over 4} \right] \quad.
\eeq
Then all intermediate results are well-defined in $\hat{\mcal{H}}$. 
One obtains
\begin{eqnarray}
\lefteqn{H(\hat{x}^{\mrm{c}}, \hat{p}^{\mrm{c}}) \xi_n} \\ 
& = & {(n - 1 + 1 / 2)^{3 / 2} \over 2} \hat{\psi}_{n - 1} 
+ {(n + 1 + 1 / 2)^{3 / 2} \over 2} \hat{\psi}_{n + 1} \nonumber 
\end{eqnarray}
for $n \geq 2$. We emphasize that this result cannot be used to check 
the eigenvalue equation for 
$\hat{\psi}_n = \lim_{N \to \infty} f^{(n)}_N$ (cf.\ (\ref{3aa})): The 
Hamiltonian is an unbounded operator, hence one cannot expect that
\beq
H \lim_{N \to \infty} f^{(n)}_N =
\lim_{N \to \infty} H f^{(n)}_N
\quad.
\eeq
In fact, $\lim_{N \to \infty} H f^{(n)}_N$ is divergent. But one can
apply $\hat{\psi}_n$ directly to (\ref{ad}). Again, all intermediate
results are well-defined, for example
\beq
\left[ \hat{x}^{\mrm{c}} \left( 1 - (\hat{p}^{\mrm{c}})^2 \right) 
+ 2 \imu \hat{p}^{\mrm{c}} \right] \hat{\psi}_n 
= {n (n + 1) \over 2 \sqrt{n + 1 / 2}} \xi_n \quad \mbox{for } 
n \geq 1 \quad,
\eeq
and one obtains the expected result 
$H \hat{\psi}_n = (n + 1 / 2)^2 \hat{\psi}_n / 2$. It even turns out 
that this result holds for $n = 0$, so that the representation 
(\ref{ad}) is correct in the entire space 
$\hat{\mcal{V}} = \vspan \{\hat{\psi}_0, \hat{\psi}_1, \ldots\}$.

\section{The connection between real spectrum and $\mcal{PT}$ 
symmetry}
\label{sb}
 
In \cite{Bender...}, Bender et~al.\ attribute the reality of the
spectrum of non-Hermitian Hamiltonians $H$ to the $\mcal{PT}$
invariance of $H$. In fact, many examples of non-Hermitian, 
$\mcal{PT}$ symmetric Hamiltonians with real spectra have subsequently 
been found \cite{Znojil1,Bagchi,Andrianov,Fernandez}. Still, also 
examples of non-Hermitian Hamiltonians with real spectra that are not 
$\mcal{PT}$ invariant are known, see, e.~g., 
\cite{Znojil2,Roy,Cannata}. A proof that $\mcal{PT}$ invariance is 
related to the reality of the spectrum does not exist. Here we want to 
present some arguments against such a relation. 

Let us consider general properties of non-Hermitian, $\mcal{PT}$ 
symmetric Hamiltonians. We put $\Theta \equiv \mcal{PT}$ in the 
following and assume
\beq
[H, \Theta] = 0 \quad.
\eeq
Since $\Theta$ is anti-unitary, we find for an $H$ eigenstate 
$\psi_{E_n}$, $H \psi_{E_n} = E_n \psi_{E_n}$ ($E_n$ being complex in 
general), 
\beq
H \Theta \psi_{E_n} = \Theta H \psi_{E_n} = \Theta E_n \psi_{E_n}
= E_n^* \Theta \psi_{E_n} \quad,
\eeq
showing firstly that $\Theta \psi_{E_n}$ is an eigenstate of $H$ with
eigenvalue $E_n^*$ and secondly that the eigenvalues of a $\Theta$
symmetric, non-Hermitian Hamiltonian always occur in complex-conjugate
pairs. Thus, if the spectrum of $H$ is non-degenerate, one has
\beq
\Theta \psi_{E_n} = \mbox{const} \, \psi_{E_n^*} \quad.
\eeq

We want to emphasize that the $\Theta$ symmetry of a non-Hermitian
Hamiltonian $H$ is {\em not\/} sufficient to ensure the reality of the
spectrum of $H$ \cite{Bender...}. Only if one makes the additional
assumption that $\psi_{E_n}$ is a simultaneous eigenstate of $\Theta$,
i.~e.\ $\Theta \psi_{E_n} = \gamma \psi_{E_n}$, one can conclude that
the spectrum of $H$ is real.

But it is important to note that the last assumption is crucial for
the reality of the spectrum, and that there is no reason to believe
that this assumption holds in general. To see this, recall the usual
argument that is used to show that a Hamiltonian $H$ and a further
{\em linear\/} operator $A$ can be simultaneously diagonalized if
$[H, A] = 0$: If $H \psi_n = E_n \psi_n$, one has
\beq\label{ba}
H A \psi_n = A H \psi_n = A E_n \psi_n = E_n A \psi_n \quad,
\eeq
showing that $A \psi_n$ is an eigenstate with {\em the same
eigenvalue\/} as $\psi_n$. If now the spectrum of $H$ is
non-degenerate, one has $A \psi_n = \mbox{const} \, \psi_n$, and if 
the spectrum of $H$ is degenerate, one may still diagonalize $A$ in
the eigenspace corresponding to $E_n$, so that the eigenstates of $H$ 
with eigenvalue $E_n$ may still be superimposed to give eigenstates of 
$A$. If, however, $A$ is an {\em anti-linear\/} operator, then 
(\ref{ba}) gives $H A \psi_n = E_n^* A \psi_n$, so that $A \psi_n$ is 
not an eigenstate with the eigenvalue $E_n$. Hence, in general one 
cannot simultaneously diagonalize $H$ and an anti-linear operator $A$, 
even if their commutator vanishes.
\end{document}